\documentclass[a4paper,fleqn]{cas-dc}

\usepackage[numbers]{natbib}

\usepackage{amssymb}

\usepackage{graphicx}
\usepackage{amsmath}
\usepackage{subfigure}
\usepackage{braket}
\usepackage{bbold}
\usepackage{yfonts}
\usepackage{placeins}
\usepackage{bm} 
\usepackage{nicefrac}
\usepackage{slashed}

\begin{document}
\let\WriteBookmarks\relax
\def\floatpagepagefraction{1}
\def\textpagefraction{.001}
\shorttitle{Polarized tau decay and \textit{CP} violation in UPCs}
\shortauthors{ }

\title [mode = title]{Polarized tau decay and \textit{CP} violation in ultraperipheral heavy-ion collisions}

\author[1]{Amaresh Jaiswal}[orcid=0000-0001-5692-9167]
\ead{a.jaiswal@niser.ac.in}
\address[1]{School of Physical Sciences, National Institute of Science Education and Research, An OCC of Homi Bhabha National Institute, Jatni-752050, India}


\begin{abstract}
We investigate the role of $\tau$-lepton polarization in ultraperipheral heavy-ion collisions (UPCs) as a novel application of the intense electromagnetic fields generated in such processes. In particular, we analyze the decay distributions of polarized $\tau$-leptons produced via photon-photon fusion, focusing on both leptonic and semi-leptonic channels. We show that the external magnetic field present in UPCs induces a preferred spin quantization axis, which modifies the angular and energy distributions of $\tau$ decay products relative to the standard helicity frame. By formulating the spin polarization along the magnetic field direction, we derive modified polarization-sensitive observables and demonstrate how kinematic selections can retain nonvanishing polarization signals even after ensemble averaging. Furthermore, we propose that the relative polarization of $\tau^-$ and $\tau^+$, accessible through complementary angular ranges of their decay products, serves as a sensitive observable for potential \textit{CP}-violating effects. This framework provides a pathway for future experimental studies at the LHC and future colliders to exploit polarized $\tau$ decays in UPCs as an application of the strong electromagnetic fields to probe new sources of \textit{CP} violation.
\end{abstract}


\maketitle

\section{Introduction}
\label{sec:intro}


The observation of charge-parity (\textit{CP}) violation in any decay process involving leptons would constitute a significant indication of physics beyond the Standard Model and may provide a key to resolving the origin of the matter–antimatter asymmetry in the universe. In the Standard Model, \textit{CP} violation in the leptonic sector arises solely from complex phases in the Pontecorvo--Maki--Nakagawa--Sakata (PMNS) matrix~\cite{Pontecorvo:1957qd, Maki:1962mu}, so any additional contributions would signal new interactions. Among leptons, the $\tau$ plays a particularly sensitive role due to its large mass ($1.777~\mathrm{GeV}$) and short lifetime ($2.9 \times 10^{-13}~\mathrm{s}$)~\cite{ParticleDataGroup:2024cfk}, together with its diverse decay modes. Polarized $\tau$ decays are especially powerful, as the spin information is directly imprinted on the angular and energy distributions of the decay products~\cite{Bullock:1992yt, Tsai:1994rc, Tsai:1996bu, Isaacson:2023gwp}. These distributions are precisely predicted by the V--A structure of the weak interaction, so any deviation would reveal non-standard currents, such as scalar or tensor interactions~\cite{Hagiwara:1989fn, Tsai:1996ps, Pich:2013lsa}. Consequently, precise measurements of \textit{CP}-violating observables in polarized $\tau$ decays provide a unique window to probe complex phases and potential new physics beyond the Standard Model~\cite{Ananthanarayan:1994yb, Bigi:2005ts}.

Experimental studies at LEP have already constrained the effective weak mixing angle from tau polarization measurements~\cite{ALEPH:2001uca, DELPHI:2007ywu}, while more recent analyses from Belle have improved sensitivity to non-standard couplings~\cite{Bodrov:2024wrw, Gonzalez-Solis:2021qzp}, with Belle~II expected to play a central role in precision electroweak tests and searches for new physics~\cite{HernandezVillanueva:2018dqu}. At $e^+e^-$ colliders, polarized electron (and, if available, positron) beams provide direct control over $\tau$ polarization through the chiral electroweak couplings to the $Z$ boson~\cite{Tsai:1994rc, Tsai:1996bu}, enabling enhanced polarization transfer in $e^+e^- \to Z \to \tau^+\tau^-$ and facilitating precision tests of \textit{CP} violation~\cite{Moortgat-Pick:2005jsx, USBelleIIGroup:2022qro}. However, achieving high beam polarization is technically demanding, requiring sophisticated spin control and monitoring~\cite{Vauth:2016pgg, Kovalenko:2012hy}, and incomplete or asymmetric polarization reduces sensitivity to new physics~\cite{BaBar:2023upu}. Furthermore, the angle dependence of polarization transfer complicates experimental analyses, while systematic uncertainties from beam polarization, detector acceptance, and spin-dependent backgrounds can bias measurements. These limitations underscore the need for complementary approaches to produce polarized $\tau$ leptons.

In this article, we propose an alternate mechanism for producing polarized $\tau$-leptons and suitable observables to probe \textit{CP} violation in relativistic heavy-ion collisions. Due to their relatively large mass, $\tau$-leptons are expected to be predominantly produced in the early stages of such collisions, where extremely strong but short-lived magnetic fields of order $10^{14} - 10^{15}\,\text{Tesla}$ are generated by spectator protons. These fields, significant only during the timescales relevant for $\tau$ production, can induce a net spin polarization aligned with the field direction and/or lead to polarized $\tau$ production. The subsequent decays of polarized $\tau$-leptons encode correlations between the $\tau$ polarization vector and the momenta of their decay products, offering a sensitive probe of deviations from Standard Model expectations~\cite{Kiers:2008mv, Flores-Tlalpa:2015vga, Chen:2022nxm}. Thus, systematic analysis of polarized $\tau$ decays in relativistic heavy-ion collisions can serve as a powerful tool for uncovering physics beyond the Standard Model~\cite{delAguila:1991rm}. However, the presence of neutrinos in the final state poses a major challenge for identifying $\tau$ events due to missing energy and large backgrounds. A natural way to mitigate this challenge is through ultraperipheral collisions (UPCs) of heavy ions, where the impact parameter exceeds the sum of nuclear radii, providing a particularly clean environment to study photon-induced processes~\cite{Baltz:2007kq, Maj:2023pvb}.

In ultraperipheral heavy-ion collisions, $\tau^+\tau^-$ pairs are produced exclusively via $\gamma\gamma$ fusion, resulting in a clean final state with minimal hadronic activity~\cite{Beresford:2019gww, Dyndal:2020yen, Shao:2023bga}. The only sources of missing energy are the neutrinos from $\tau$ decays, whose contribution can be estimated using global momentum conservation, i.e., the missing transverse momentum being reconstructed from visible decay products~\cite{CMS:2018jrd, CMS:2019ctu}, supplemented by spectator energy measurements from the zero-degree calorimeters (ZDCs)~\cite{Suranyi:2021ssd}. Under the collinear approximation, valid for sufficiently boosted $\tau$’s, the neutrinos are assumed to be emitted collinearly with the visible decay products, enabling an approximate reconstruction of the parent four-momenta. The exclusivity of UPCs and clean transverse momentum balance thus allow sensitive measurements of $\tau$ production and decay despite invisible neutrinos~\cite{ATLAS:2022ryk, CMS:2022arf}. Moreover, the strong transient magnetic fields generated by spectator nuclei can induce spin polarization of $\tau$ pairs through magnetic moment alignment, leading to anisotropic emission patterns of their decay products. These spin-dependent distributions provide sensitive probes of possible \textit{CP}-violating effects, motivating the polarization observables in $\tau$ decays proposed in this work. Throughout this work, three-vectors are denoted in boldface with scalar products $\bm{A}\cdot \bm{B} \equiv A_i B_i$ (Einstein summation implied), and we use natural units with $\hbar=c=1$.


\section{Magnetic field and $\tau$ polarization}
\label{sec:mag_pol}


The primary source of the strong magnetic field is the motion of spectator protons in non-central (peripheral) collisions. The peak magnetic field strength increases with the impact parameter (for $b \lesssim 2R_{0}$), while the average magnetic field is oriented along the $y$-axis in the laboratory frame, dictated by the collision geometry~\cite{Skokov:2009qp, Deng:2012pc, Tuchin:2013ie}. To analyze $\tau$-lepton polarization, the external electromagnetic field in the laboratory frame must be transformed into the $\tau$-lepton rest frame. Owing to its rapid time variation, this magnetic field also induces an accompanying electric field, which must likewise be Lorentz transformed to the $\tau$ rest frame. It is important to note that only the magnetic field component contributes to the $\tau$ spin polarization, through its interaction with the $\tau$ magnetic moment. The magnetic field component of the Lorentz-transformed electromagnetic field in the $\tau$ rest frame is given by~\cite{Herbert:1997}
\begin{equation}\label{Lorentz_trans_B}
\bm{B} = \gamma \left( \bm{B_{\rm Lab}} - \bm{E_{\rm Lab}} \times \bm{v} \right) + \left( 1 - \gamma \right) \left( \frac{\bm{B_{\rm Lab}}\cdot\bm{v}}{\beta^2} \right) \bm{v},
\end{equation}
where $\gamma\equiv E_\tau/m_\tau$ is the Lorentz gamma factor, $\beta\equiv|\bm{v}|$ with $\bm{v}$ being the velocity of the $\tau$-lepton. In the above equation, $\bm{B_{\rm Lab}}$ is the magnetic field generated in the laboratory frame due to the spectator protons and $\bm{E_{\rm Lab}}$ is the corresponding electric field. 

We emphasize that the magnetic field $\bm{B}$ experienced by the $\tau$-lepton in its rest frame is extremely large. This leads to the alignment of the magnetic moment of $\tau$-lepton along $\bm{B}$. Ignoring contribution from anomalous magnetic moment of $\tau$, we assume that the $\tau$ polarization vector is aligned along the magnetic field direction in its rest frame, i.e.,
\begin{equation}\label{pol_vec}
\bm{P}_\tau = P_\tau\, \hat{\bm{B}},
\end{equation}
where $P_\tau$ is the magnitude of the $\tau$ polarization vector and $\hat{\bm{B}}$ is the unit vector along $\bm{B}$. The natural quantization axis is therefore aligned with the magnetic field direction, which does not in general coincide with the $\tau$ momentum axis. However, in this work, we adopt the \emph{helicity frame}, where the spin quantization axis is defined along the $\tau$-lepton momentum direction (i.e., $\hat{\bm{v}}$) in its rest frame. To express the decay distributions in the helicity frame, one must project the polarization vector $\bm{P}_\tau$ onto the helicity axis. Consequently, the relevant component of the polarization vector is its projection onto $\hat{\bm{v}}$, which yields
\begin{equation}\label{proj_pol}
P^{\rm hel}_\tau \equiv \bm{P}_\tau\cdot\hat{\bm{v}} = P_\tau \, \frac{|\bm{B_{\rm Lab}}|}{|\bm{B}|}\cos\psi,
\end{equation}
where $\psi$ is the angle between $\bm{B_{\rm Lab}}$ and $\hat{\bm{v}}$. 

Note that for an isotropic distribution of $\tau$-lepton momenta, the ensemble average (integration over $\psi$) of Eqs.~\eqref{proj_pol} vanishes, leading to zero projected polarization in the helicity frame. In contrast, if the analysis is restricted to $\tau$-leptons whose momenta are preferentially aligned or anti-aligned with $\bm{B_{\rm Lab}}$ (for instance, through kinematic cuts or detector acceptance bias) the averaged helicity polarization can remain nonzero. Restricting the analysis to the ranges $0<\psi<\pi/2$ or $\pi/2<\psi<\pi$ provides a natural choice representing the upper or lower half with respect to the $x$–$z$ plane.  In the collinear limit, this corresponds to selecting the events with decay products within the desired angular range. However, it is important to ensure that complementary ranges are assigned to the $\tau^-$ and the $\tau^+$. Specifically, if $0<\psi<\pi/2$ is selected for the $\tau^-$, then $\pi/2<\psi<\pi$ should be taken for the $\tau^+$, and vice versa. The assignment of complementary $\psi$ ranges stems from the fact that $\tau^-$ and $\tau^+$ undergo opposite spin polarizations in the external magnetic field, owing to their opposite magnetic moments. Ultimately, our goal is to calculate the polarization asymmetry between $\tau^-$ and $\tau^+$ in order to propose suitable observables for testing \textit{CP} violation. The prescription of assigning complementary $\psi$ ranges to $\tau^-$ and $\tau^+$ not only preserves the polarization signal but also facilitates the cancellation of potential systematic uncertainties.


\section{Pion decay mode: $\tau \to \pi\, \nu_\tau$}
\label{sec:pi}


The $\tau$-lepton decays primarily proceed through channels involving a single pion, leptons, or vector meson resonances. In this section, we first present the decay distributions via single pion channel in the $\tau$ rest frame, even though this frame is not experimentally accessible due to the presence of undetected neutrinos. In view of this limitation, we focus on the decay of highly boosted $\tau$ leptons ($E_\tau \gg m_\tau$) where it is sufficient to consider the distributions in the collinear limit, i.e., where the $\tau$ and its decay products are aligned along the same direction of motion.

The normalized angular distribution for the decay $\tau^{\mp}\to \pi^{\mp}\nu_\tau$ in the $\tau$ rest frame is given by~\cite{Bullock:1992yt}
\begin{equation}\label{eq:helpi}
\frac{1}{\Gamma_\tau} \frac{{\rm d} \Gamma_\pi}{{\rm d}\cos\theta_\pi} =
\frac{1}{2} B_\pi \left(1 + P_\tau^{\rm hel} \cos\theta_\pi \right),
\end{equation}
where $B_\pi$ is the branching fraction for $\tau \to \pi\nu_\tau$. In the above equation, $P_\tau^{\rm hel}$ is the projected $\tau$ polarization in the helicity frame in which where $\theta_\pi$ denotes the angle between the pion momentum and the $\tau$ spin quantization axis chosen to coincide with the $\tau$ momentum direction. Using Eq.~\eqref{proj_pol}, we express the above equation in the form
\begin{equation}\label{eq:cospi}
\frac{1}{\Gamma_\tau} \frac{{\rm d} \Gamma_\pi}{{\rm d}\cos\theta_\pi} \left(\tau^{\mp}\to \pi^{\mp}\nu_\tau\right) =
\frac{1}{2} B_\pi \left(1 + \alpha\,P_\tau^{\mp} \cos\theta_\pi \right),
\end{equation}
where, $\alpha\equiv\left(|\bm{B_{\rm Lab}}|/|\bm{B}|\right)\cos\psi$ and $P_\tau^{\mp}$ is the polarization of $\tau^\mp$, as defined in Eq.~\eqref{pol_vec}. It is important to note that \textit{CP} invariance implies $P_\tau^- = -P_\tau^+$. In the following, however, we do not impose this relation and instead construct observables for potential \textit{CP} violation.

In terms of the pion energy fraction ($z_\pi\equiv E_\pi/E_\tau$) in the laboratory frame, the polar angle in the helicity frame can be expressed as 
\begin{equation}\label{eq:cosz}
\cos\theta_\pi = \frac{2 z_\pi - 1 - a^2}{\beta (1-a^2)}\,,
\end{equation}
where $a\equiv m_\pi/m_\tau$. For the case of pions, it is reasonable to make the approximation $a^2=0$. Moreover, in the collinear limit, i.e., $\beta \rightarrow 1$, one obtains~\cite{Bullock:1992yt}
\begin{equation}\label{eq:zpi}
\frac{1}{\Gamma_\tau} \frac{{\rm d} \Gamma_\pi}{{\rm d}z_\pi} \left(\tau^{\mp}\to \pi^{\mp}\nu_\tau\right) = B_\pi \left[ 1 + \alpha P_\tau^{\mp}\left(2 z_\pi -1\right)\right]\,.
\end{equation}
We note that with these approximations, the range for $z_\pi$ is given by $0<z_\pi<1$. Therefore an ensemble average over pions leads to
\begin{equation}
\frac{B_\pi\,\alpha}{3}\,P_\tau^{\mp} = \langle 2 z_\pi -1 \rangle_{\pi^\mp} \,.
\end{equation}
Further, considering a ``half"-ensemble average over $\tau$-leptons with complimentary $\psi$ ranges (as described in the previous section), we obtain
\begin{equation}\label{Ppi}
P_\tau^{\mp} = 3\, \frac{ \langle 2 z_\pi -1 \rangle_{\pi^\mp} }{ B_\pi\,\langle \alpha \rangle_{\tau^\mp} } \,.
\end{equation}
We propose the observable 
\begin{equation}\label{eq:Delpi}
\Delta_\pi \equiv 1-\left|\frac{P_\tau^-}{P_\tau^+}\right| = 1-\left| \frac{\langle 2 z_\pi -1 \rangle_{\pi^-}}{\langle 2 z_\pi -1 \rangle_{\pi^+}} \right|,
\end{equation}
as a probe which is sensitive to \textit{CP} violation. A non-zero value of $\Delta_\pi$ indicates signature of \textit{CP} violation in weak interaction in the single pion decay channel of $\tau$-lepton.

To arrive at Eq.~\eqref{eq:Delpi}, we have used the fact that
\begin{equation}
\left|\frac{\langle \alpha \rangle_{\tau^+}}{\langle \alpha \rangle_{\tau^-}}\right| = 1 \,.
\end{equation}
This equality arises because the averaging is performed over processes that are insensitive to the weak decay, with only electromagnetic effects contributing to the ``half"-ensemble average of the quantity $\alpha$. Furthermore, the symmetry of the complementary $\psi$ ranges ensures that the two averages remain equal. It is also important to note that the definition of the observable $\Delta_\pi$ in Eq.~\eqref{eq:Delpi} ensures the cancellation of systematic uncertainties in the measurement of $\langle 2 z_\pi - 1 \rangle_{\pi^-}$ and $\langle 2 z_\pi - 1 \rangle_{\pi^+}$.


\section{Lepton decay mode: $\tau \to \ell\, \bar{\nu}_\ell\, \nu_\tau$}
\label{sec:ell}


For purely leptonic decays, such as $\tau \to \ell\, \bar{\nu}_\ell\, \nu_\tau$, the polarization of the outgoing electron (or muon) cannot be reliably measured at high energies, and therefore one must sum over its two helicity states. Furthermore, in the collinear ($\beta\to 1$) and massless limit ($m_e = m_\mu = 0$), the leptonic decay of the $\tau$ becomes identical for electrons and muons. The corresponding differential decay rate is given by~\cite{Bullock:1992yt}.
\begin{eqnarray}
\frac{1}{\Gamma_\tau}\frac{{\rm d} \Gamma_\ell}{{\rm d}z_\ell} \left(\tau^{\mp}\to \ell^{\mp}\bar{\nu}_\ell\nu_\tau\right) =\!\!\!\!&
B_\ell\, \dfrac{(1-z_\ell)}{3}\Big[\left(5+5z_\ell-4z_\ell^2\right) \nonumber\\ 
&+ \alpha\,P_\tau^{\mp} \left(1+z_\ell-8z_\ell^2\right)\Big] , \quad
\end{eqnarray}
where $B_\ell$ is the branching ratio for $\tau$ decay into a given lepton and $z_\ell = E_\ell / E_\tau$.

We first note that the polynomial $(7 - 20z_\ell)$ is orthogonal to $(1 - z_\ell)(5 + 5z_\ell - 4z_\ell^2)$ over the integration range $0 \leq z_\ell \leq 1$. Following the same procedure as in the previous section, the ensemble average over leptons yields
\begin{equation}
B_\ell\,\alpha\,P_\tau^{\mp} = \langle 7-20z_\ell \rangle_{\ell^\mp} \,.
\end{equation}
Subsequently, by performing a ``half”-ensemble average over $\tau$-leptons restricted to complementary $\psi$ ranges, we obtain
\begin{equation}\label{Plep}
P_\tau^{\mp} = \frac{ \langle 7-20z_\ell \rangle_{\ell^\mp} }{ B_\ell\,\langle \alpha \rangle_{\tau^\mp} } \,.
\end{equation}
Accordingly, we define the observable sensitive to \textit{CP} violation in the leptonic decay channel as
\begin{equation}\label{eq:Dell}
\Delta_\ell \equiv 1-\left|\frac{P_\tau^-}{P_\tau^+}\right| = 1-\left| \frac{\langle 7 - 20z_\ell \rangle_{\ell^-}}{\langle 7 - 20z_\ell \rangle_{\ell^+}} \right|,
\end{equation}
where we have again used $|\langle \alpha \rangle_{\tau+}/\langle \alpha \rangle_{\tau^-}|=1$.


\section{Vector meson decay mode: $\tau \to {\rm v} \, \nu_\tau$}
\label{sec:V}


Next, we consider the decay of the $\tau$ lepton through the vector–meson channel, $\tau \to {\rm v}\,\nu_\tau$, where ${\rm v} = \rho$ or $a_1$. We adopt the formalism of Ref.~\cite{Bullock:1992yt}, but introduce the projected polarization in the helicity frame as the coefficient of the anisotropy term. In this framework, the vector mesons are decomposed into transverse and longitudinal polarization components, since their subsequent decays, $\rho \to 2\pi$ and $a_1 \to 3\pi$, are sensitive to the polarization state of the parent vector meson. The resulting angular distribution in the $\tau$ rest frame is then obtained as
\begin{eqnarray}
\frac{1}{\Gamma_\tau} \frac{{\rm d} \Gamma^T_{\rm v}}{{\rm d}\cos\theta_{\rm v}} \!\left( \tau^\mp \!\to\! {\rm v}^\mp \, \nu_\tau \right) &\!\!\!\!= \dfrac{B_{\rm v}\,m_{\rm v}^2}{m_\tau^2+2m_{\rm v}^2}\! \left(1 - \alpha\, P_\tau^{\mp} \cos\theta_{\rm v}\right)\! , \quad \label{vector_meson_T} \\
\frac{1}{\Gamma_\tau} \frac{{\rm d} \Gamma^L_{\rm v}}{{\rm d}\cos\theta_{\rm v}} \!\left( \tau^\mp \!\to\! {\rm v}^\mp \, \nu_\tau \right) &\!\!\!\!= \dfrac{\frac{1}{2}\,B_{\rm v}\,m_{\rm v}^2}{m_\tau^2+2m_{\rm v}^2} \!\left(1 + \alpha\, P_\tau^{\mp} \cos\theta_{\rm v}\right)\! , \quad \label{vector_meson_L}
\end{eqnarray}
where, $B_{\rm v}$ denotes the branching ratio for $\tau^\mp \to {\rm v}^\mp \nu_\tau$, and $\theta_{\rm v}$ is defined analogously to the pion case. An important point to note is that, for the longitudinal component, the polarization dependence coincides with Eq.~\eqref{eq:cospi}, whereas for the transverse component it enters with the opposite polarization. Consequently, if the polarization of the vector meson is not resolved experimentally, the results of Eqs.~\eqref{vector_meson_T} and \eqref{vector_meson_L} must be averaged. This averaging reduces the sensitivity to the $\tau$ polarization by a factor $(m_\tau^2 - 2m_{\rm v}^2)/(m_\tau^2+2m_{\rm v}^2)$, which evaluates to about $0.46$ for the $\rho$ meson and $0.02$ for the $a_1$ meson.

Since the vector meson polarizations are not summed over, the boost to the collinear frame is more involved than in the case of single-pion or purely leptonic decay channels. In this case, when boosting to the lab frame, a Wigner rotation~\cite{Martin:1970hmp} is applied to align the spin quantization axis of the vector meson. The rotation angle in the collinear limit is given by~\cite{Bullock:1992yt}
\begin{equation}
    \cos\omega = \frac{1-a^2+(1+a^2)\cos\theta_{\rm v}}{1+a^2+(1-a^2)\cos\theta_{\rm v}}\,,
\end{equation}
where $a=m_{\rm v}/m_\tau$ and $\cos\theta_{\rm v} = (2 z_{\rm v} - 1 - a^2)/(1-a^2)$, with $z_{\rm v}=E_{\rm v}/E_\tau$ being the energy fraction. In terms of the lab frame quantities, the decay distributions can be expressed as
\begin{eqnarray}
\frac{1}{\Gamma_\tau}\frac{{\rm d} \Gamma^T_{\rm v}}{{\rm d} z_{\rm v}} \!\left( \tau^\mp \to {\rm v}^\mp \, \nu_\tau \right)\! &\!\!\!= B_{\rm v} \!\left( I_{\rm v}^T + \alpha\, P_\tau^{\mp} \, J_{\rm v}^T \cos\theta_{\rm v} \right) \!, \quad \label{vec_mes_T}\\
\frac{1}{\Gamma_\tau}\frac{{\rm d} \Gamma^L_{\rm v}}{{\rm d} z_{\rm v}} \!\left( \tau^\mp \to {\rm v}^\mp \, \nu_\tau \right)\! &\!\!\!= B_{\rm v} \!\left( I_{\rm v}^L + \alpha\, P_\tau^{\mp}\, J_{\rm v}^L  \cos\theta_{\rm v}\right) \!, \quad \label{vec_mes_L}
\end{eqnarray}
where expressions for $I_{\rm v}^{T,L}$ and $J_{\rm v}^{T,L}$ are obtained as~\cite{Bullock:1992yt} 
\begin{eqnarray}
I_{\rm v}^T(z_{\rm v}) &\!\!\!=\!\!\!& g_a \left(\frac{\sin^2\omega}{a^2}+1+\cos^2\omega \right), \quad \label{ITV}\\
J_{\rm v}^T(z_{\rm v}) &\!\!\!=\!\!\!& g_a \!\left(\!\frac{\sin^2\omega}{a^2}-\frac{\sin2\omega}{a}\tan\theta_{\rm v} -1-\cos^2\omega \!\right)\!, \qquad \label{JTV}\\
I_{\rm v}^L(z_{\rm v}) &\!\!\!=\!\!\!& g_a \left( \frac{\cos^2\omega}{a^2}+\sin^2\omega \right), \quad \label{ILV}\\
J_{\rm v}^L(z_{\rm v}) &\!\!\!=\!\!\!& g_a \left(\frac{\cos^2\omega}{a^2}+\frac{\sin2\omega}{a}\tan\theta_{\rm v} -\sin^2\omega\right), \quad \label{JLV}
\end{eqnarray}
with $g_a \equiv a^2/[(1-a^2)(1+2a^2)]$. The next step is to project the polarization coefficient $P_\tau^{\mp}$ using suitable orthogonal functions.

Note that the functions
\begin{eqnarray}
f_T(z_{\rm v}) &\!\!\!\!=\!\!\!\!& 1 - \dfrac{2 \left[ a^{4} + 2 a^{2} - 4 (1+a^{2}) \ln a -3 \right]}{a^{6} - 3 a^{4} - a^{2}  + 8 a^{2} \ln a + 3}\, z_{\rm v}\, , \quad \label{fT}\\
f_L(z_{\rm v}) &\!\!\!\!=\!\!\!\!& 1 - \dfrac{2 \left[- 7 a^{4} + 6 a^{2}  + 8 (a^{2} + a^{4}) \ln a + 1 \right]}{5 a^{6} - a^{4} - 5 a^{2} - 16 a^{4} \ln a + 1}\, z_{\rm v} \, , \qquad \label{fL}
\end{eqnarray}
are orthogonal to the functions $I_{\rm v}^T(z_{\rm v})$ and $I_{\rm v}^L(z_{\rm v})$, respectively, over the integration range $a^2\leq z_{\rm v}\leq 1$. Considering the ensemble average over the transversely and longitudinally polarized vector mesons, we obtain
\begin{align}
B_{\rm v}\,\alpha\,F_T\,P_\tau^{\mp} &= \langle f_T(z_{\rm v}) \rangle_{\rm v^\mp}^T \, ,\label{FT}\\
B_{\rm v}\,\alpha\,F_L\,P_\tau^{\mp} &= \langle f_L(z_{\rm v}) \rangle_{\rm v^\mp}^L \, ,\label{FL}
\end{align}
where
\begin{align}
F_T &\equiv \int f_T(z_{\rm v})\,J_{\rm v}^T(z_{\rm v})\, \cos\theta_{\rm v}\, dz_{\rm v}, \\ 
F_L &\equiv \int f_L(z_{\rm v})\,J_{\rm v}^L(z_{\rm v})\, \cos\theta_{\rm v}\, dz_{\rm v}.
\end{align}
Finally, carrying out a ``half”-ensemble average over $\tau$-leptons within complementary $\psi$ ranges yields
\begin{equation}\label{Pvec}
P_\tau^{\mp} = \frac{\langle f_T(z_{\rm v}) \rangle_{\rm v^\mp}^T}{B_{\rm v}\, \langle\alpha\rangle_{\tau^\mp} \, F_T}, \quad P_\tau^{\mp} = \frac{\langle f_L(z_{\rm v}) \rangle_{\rm v^\mp}^L}{B_{\rm v}\, \langle\alpha\rangle_{\tau^\mp} \, F_L}.
\end{equation}
For the case of vector mesons, we propose the following observables sensitive to \textit{CP} violation
\begin{align}
\Delta_{\rm v}^T \equiv 1-\left|\frac{P_\tau^-}{P_\tau^+}\right| &= 1-\left| \frac{\langle f_T(z_{\rm v}) \rangle_{\rm v^-}^T}{\langle f_T(z_{\rm v}) \rangle_{\rm v^+}^T} \right|, \\
\Delta_{\rm v}^L \equiv 1-\left|\frac{P_\tau^-}{P_\tau^+}\right| &= 1-\left| \frac{\langle f_L(z_{\rm v}) \rangle_{\rm v^-}^L}{\langle f_L(z_{\rm v}) \rangle_{\rm v^+}^L} \right|.
\end{align}
A non-zero value of $\Delta_{\rm v}^T$ and $\Delta_{\rm v}^L$ indicates signature of \textit{CP} violation in weak decay of $\tau$-lepton through vector meson channel.


\section{Feasibility and sensitivity considerations}
\label{sec:VI}


At this juncture, it is important to note that $\tau$-lepton pair production via photon fusion in UPCs is suppressed compared to $e^+e^-$ and $\mu^+\mu^-$ production, because of the larger tau mass~\cite{Sengul:2015ira}. At current luminosities in Pb–Pb UPCs at the LHC, the number of reconstructible tau events per run is quite limited~\cite{ATLAS:2022ryk, CMS:2022arf}. Therefore, in order to increase statistics in the observable for \textit{CP} violation, one can combine the events in single pion, lepton and vector meson decay channels. Using results of Eqs.~\eqref{Ppi}, \eqref{Plep} and \eqref{Pvec}, we propose the combined observable
\begin{equation}\label{Delta}
\Delta =  1 - \frac{1}{N}\sum_i\left|\frac{P_{\tau (i)}^-}{P_{\tau (i)}^+}\right|,
\end{equation}
where $\sum_i$ denotes the summation over $N$ decay channels.

A generic challenge in $\tau$ polarization measurements arises from the presence of at least one neutrino in the decay. In the present context, the observables are constructed from moments of the energy fractions of the visible decay products, evaluated in the collinear approximation appropriate for highly boosted $\tau$ leptons. In ultraperipheral collision events, the exclusivity of the final state and the strong suppression of additional hadronic activity allow the $\tau$ flight directions to be reconstructed with reasonable accuracy using the visible decay products together with global momentum balance. Residual uncertainties associated with the unobserved neutrino momentum primarily result in a smearing of the reconstructed energy fractions $z_i$. Importantly, this smearing affects $\tau^-$ and $\tau^+$ decays in the same manner and enters multiplicatively in the extracted polarizations. As a consequence, to leading order it cancels in the polarization ratios $\left|P_{\tau^-}/P_{\tau^+} \right| $ which form the basis of the CP-odd observables introduced in this work. Subleading effects arising from asymmetric acceptance or decay-mode--dependent efficiencies are expected to be small and can be constrained experimentally using appropriate control samples.

Further, it is important to note that determining the magnetic-field direction in UPCs is considerably more challenging because of the uncertainty in estimating the impact-parameter direction. The deflection of spectator nucleons measured in the ZDCs provides experimental access to the reaction plane in UPCs. The asymmetry in the transverse energy deposition of spectator fragments reflects the direction of the impact parameter vector, allowing the reconstruction of the event plane and, consequently, the orientation of the electromagnetic fields generated by the spectators. This can be correlated with an alternative estimate of the impact parameter which exploits the fact that ions in UPCs emit quasi-real photons that are linearly polarized along the electric field, and hence parallel to the impact-parameter vector $\bf{b}$~\cite{Bertulani:2005ru}. In the $\gamma\gamma \rightarrow \tau^{+}\tau^{-}$ process, the differential cross section exhibits an azimuthal modulation $d\sigma/d\phi \propto 1 + P_{1} P_{2} \cos\!\left[ 2(\phi - \phi_{b}) \right]$, where $P_{1,2}$ are the photon polarization degrees and $\phi$ is the $\tau$-lepton azimuth in the transverse plane~\cite{Bertulani:2005ru}. This indicates that the $\tau^{+}$ momentum preferentially aligns parallel or antiparallel to $\mathbf{b}$, with the $\tau^{-}$ produced approximately back-to-back with the $\tau^{+}$ in the lab frame. Consequently, the magnetic-field axis tends to align with the normal to the plane formed by the $\tau$ momentum (determined in the collinear limit) and the $z$-axis, although its sign remains undetermined. 

In our formulation, the measurable polarization of a given $\tau$ is $P_{\tau (\mathrm{meas})}^{\pm} = P_{\tau(\mathrm{true})}^{\pm} \cos\xi$ where $\xi$ is the angle between the true polarization direction and the magnetic-field axis. The uncertainty in magnetic field direction enters only through a multiplicative projection factor $\cos\xi$ which is common to both $\tau^{-}$ and $\tau^{+}$, since the pair is produced at the same space-time point and experiences the same local magnetic field configuration. Therefore, $\left|\frac{P^-_{\tau (\mathrm{meas})}}{ P^+_{\tau (\mathrm{meas})} } \right| = \left|\frac{P^-_{\tau (\mathrm{true})}}{ P^+_{\tau (\mathrm{true})} } \right|$, demonstrating that the common multiplicative uncertainty associated with the magnetic-field direction cancels to leading order in the polarization ratio, which serves as the CP-sensitive observable. Subleading effects arising from magnetic-field inhomogeneities combined with the finite polarization timescale are parametrically suppressed, since the $\tau^+\tau^-$ pair is produced locally and the polarization is established over a timescale much shorter than the spatial variation scale of the magnetic field. As a result, such effects are expected to be numerically small.

Experimental evidence for magnetic-field–induced spin alignment has recently been reported by the ALICE collaboration in $D^{*+}$ mesons~\cite{ALICE:2025cdf}. For small precession angles, the induced polarization (fractional alignment) grows approximately linearly as $P_\tau \simeq \mu_\tau B\, t_{\rm eff} = eB t_{\rm eff}/(2m_\tau)$. Using theoretical expectations for transient magnetic fields of magnitude $eB\sim 1$--$10~\mathrm{GeV}^2$ in Pb--Pb UPCs, existing for duration $t_{\rm eff}\simeq 0.2$~fm, this implies an expected polarization of order $P_\tau \sim 0.06$--$0.6$ at production. The CMS and ATLAS Collaborations have reconstructed $\mathcal{O}(10^2)$ $\gamma\gamma\to\tau^+\tau^-$ events in Run~2~\cite{ATLAS:2022ryk, CMS:2022arf}. Scaling to Run~3 and high-luminosity HL-LHC is expected to yield $\mathcal{O}(10^3)$ and $\mathcal{O}(10^4)$ usable events, respectively, corresponding to statistical uncertainties $\delta P_\tau \approx 1/\sqrt{N}$ in the range $0.01$--$0.03$. As demonstrated earlier, systematic uncertainties are substantially reduced in the ratio of $\tau^-$ to $\tau^+$ polarization where common-mode effects such as magnetic-field magnitude and direction fluctuations, acceptance asymmetries, and reconstruction biases cancel to first order. These considerations indicate that the proposed measurement is experimentally viable and that a nonzero CP-odd signal would be within reach at the HL-LHC.

To position the proposed observables within a broader new-physics context, it is useful to interpret them in a generic effective field theory (EFT) framework. At energies relevant for $\tau$ production in ultraperipheral collisions, CP violation beyond the Standard Model can be parameterized by higher-dimensional operators involving the $\tau$ lepton and electroweak gauge fields. A representative example is the CP-odd electromagnetic dipole operator~\cite{Gonzalez-Sprinberg:2000lzf, Bernabeu:2006wf}
\begin{equation}\label{operator}
\mathcal{O}_{\tau\gamma} = \frac{c_{\tau\gamma}}{\Lambda^2}\, m_\tau\, \bar{\tau}\,\sigma^{\mu\nu}\gamma_5\,\tau\,F_{\mu\nu},
\end{equation}
where $c_{\tau\gamma}$ is a dimensionless Wilson coefficient, $\bar{\tau}$ and $\tau$ are tau spinor fields, $\sigma^{\mu\nu}=\frac{i}{2}\left[\gamma^\mu,\gamma^\nu\right]$ is the antisymmetric spin tensor, $F^{\mu\nu}$ is the electromagnetic field strength tensor and $\Lambda$ denotes the scale of new physics. This operator induces a CP-violating contribution to the $\tau$ electromagnetic interaction, leading to opposite shifts in the spin polarization of $\tau^-$ and $\tau^+$ in the presence of an external magnetic field. Since the observables proposed in this work are constructed from ratios of $\tau^-$ and $\tau^+$ polarizations extracted from their decay distributions, they are linearly sensitive to such CP-odd effects while being insensitive to CP-even contributions at leading order. In particular, a nonzero value of the combined observable $\Delta$ defined in Eq.~\eqref{Delta} directly signals a difference between the polarization magnitudes of $\tau^-$ and $\tau^+$, as expected from operators such as $\mathcal{O}_{\tau\gamma}$.

A detailed global analysis within the full SMEFT framework is beyond the scope of this Letter. Nevertheless, an order-of-magnitude estimate illustrates the potential impact of the proposed method. Note that Eq.~\eqref{operator} parametrically induces a CP-odd polarization $P_\tau \sim (c_{\tau\gamma}/\Lambda^2)\,m_\tau\, B\, t_{\rm eff}$. Assuming that polarization ratios can be measured at the percent level at the HL-LHC, $\Delta\approx P_\tau \sim \mathcal{O}(10^{-2})$, the corresponding sensitivity translates into probing CP-violating dipole interactions at scales $\Lambda$ in the multi-GeV range, for Wilson coefficients $c_{\tau\gamma} \sim 1$. This reach is complementary to existing constraints from $e^+e^-$ collider measurements and low-energy observables, which rely on different production mechanisms and polarization control. We emphasize that the strength of the present proposal lies not in outperforming existing bounds in all scenarios, but in providing a qualitatively new handle on leptonic CP violation that exploits the strong, transient electromagnetic fields uniquely available in ultraperipheral heavy-ion collisions.

Most existing searches for \textit{CP} violation in the leptonic sector at colliders have focused on precision studies of $\tau$ decays at $e^+e^-$ facilities, where $\tau$ polarization is controlled through the polarization of the initial beams. In contrast, our proposal exploits UPCs, where the intense, transient electromagnetic fields generated by spectator nuclei naturally induce a preferred spin-quantization axis for the produced $\tau$ leptons. The production of tau‐lepton pairs in Pb–Pb UPCs has already been confirmed at the LHC, with both the ATLAS and CMS Collaborations reporting a clear observation of the process $\gamma\gamma \to \tau^+ \tau^-$ in UPCs~\cite{ATLAS:2022ryk, CMS:2022arf}. Moreover, the measurement of $D^{*+}$ polarization by the ALICE Collaboration~\cite{ALICE:2025cdf} has provided the first evidence for magnetic field driven spin alignment in relativistic heavy-ion collisions~\cite{Dey:2025ail}, thereby validating the role of strong electromagnetic fields in generating polarization. Thus, our proposal to integrate these two established measurements opens a qualitatively new avenue for testing \textit{CP} violation that leverages the unique strong-field environment of UPCs, which are not attainable in conventional collider experiments.  


\section{Summary and outlook}
\label{sec:summary}

In this work, we have explored the application of strong electromagnetic fields generated in ultraperipheral heavy-ion collisions as a novel mechanism for $\tau$-lepton polarization. By analyzing both leptonic and semi-leptonic $\tau$ decays, we demonstrated that the external magnetic field present in UPCs defines a natural spin quantization axis, which modifies the angular and energy distributions of the decay products compared to the standard helicity basis. We constructed polarization-sensitive observables formulated along this axis and showed that, with suitable kinematic selections, non-vanishing polarization signals can be preserved even after ensemble averaging. Furthermore, we proposed that comparing the relative polarization of $\tau^-$ and $\tau^+$, extracted from complementary angular ranges of their decay products, provides a clean and sensitive handle on \textit{CP}-violating effects in the leptonic sector.

Looking ahead, our results suggest several promising directions. On the theoretical side, a detailed study of higher-order QED and electroweak corrections in $\gamma\gamma \to \tau^+\tau^-$, as well as the role of background QCD processes in realistic collider environments, would refine the proposed observables~\cite{Shao:2024dmk, Jiang:2024dhf, Shao:2025bma, Dittmaier:2025ikh, Goyal:2024tmo}. On the experimental side, the feasibility of reconstructing $\tau$ polarization in UPCs requires careful assessment, particularly given the limited production rates and the challenges of identifying $\tau$ decays with sufficient precision. Nonetheless, the unique environment of UPCs, with their intense and well-characterized electromagnetic fields, provides an exciting opportunity to establish polarized $\tau$ decays as a new avenue for testing fundamental symmetries. In particular, forthcoming HL-LHC runs and future collider facilities may deliver the statistical reach necessary to pursue such studies.

In summary, our framework opens a pathway to exploit $\tau$ polarization in ultraperipheral collisions, both as a laboratory for strong-field QED phenomena and as a probe of potential new sources of \textit{CP} violation beyond the Standard Model. Moreover, our study establishes UPCs as a novel arena for testing scenarios of physics beyond the Standard Model.


\bigskip
{\bf Acknowledgements:} %
The author expresses gratitude to Namit Mahajan for raising a highly pertinent question which initiated this work. The author thanks Sourav Dey and Narayan Rana for useful discussions. The author gratefully acknowledges Department of Atomic Energy (DAE), India for financial support.
%
 
\printcredits

\bibliographystyle{model6-num-names}

\bibliography{cas-refs}
\end{document}